\documentclass[runningheads]{llncs}
\usepackage[T1]{fontenc}
\usepackage{graphicx}
\usepackage{array}
\usepackage{braket}
\usepackage{booktabs}
\usepackage{amsmath,amssymb,amsfonts}
\usepackage{hyperref}

\begin{document}
\title{Variational Quantum Algorithms for Particle Track Reconstruction}
\titlerunning{VQA for Particle Track Reconstruction}
%
\author{
Vincenzo Lipardi$^1$ \textsuperscript{\textdagger} \orcidID{0009-0000-3243-7969} \and  Xenofon Chiotopoulos$^{1,2,3}$ \textsuperscript{\textdagger} \orcidID{0009-0006-5762-6559}
\and  Jacco A. de Vries$^{2,3}$ \orcidID{0000-0003-4712-9816} 
\and  Domenica Dibenedetto$^1$ \orcidID{0000-0002-2538-3170} 
\and
Kurt Driessens$^1$ \orcidID{0000-0001-7871-2495} \and Marcel Merk $^{2,3}$ \orcidID{0000-0003-0818-4695} \and Mark H.M. Winands$^1$\orcidID{0000-0002-0125-0824}}
\authorrunning{Lipardi et al.}
%
\institute{Department of Advanced Computing Sciences, \\ Maastricht University, The Netherlands  \\
\and 
Gravitational Waves and Fundamental Physics, \\ Maastricht University, The Netherlands \\
\email{\{vincenzo.lipardi, xenofon.chiotopoulos\}@maastrichtuniversity.nl}
\and Nikhef National Institute for Subatomic Physics, Amsterdam, The Netherlands }
%
\maketitle              
\def\thefootnote{\textdagger}\footnotetext{These authors contributed equally to this work. The order of the first two authors has been chosen by flipping a quantum coin.}\def\thefootnote{\arabic{footnote}}

\begin{abstract}
Quantum Computing is a rapidly developing field with the potential to tackle the increasing computational challenges faced in high-energy physics. In this work, we explore the potential and limitations of variational quantum algorithms in solving the particle track reconstruction problem. We present an analysis of two distinct formulations for identifying straight-line tracks in a multilayer detection system, inspired by the LHCb vertex detector. The first approach is formulated as a ground-state energy problem, while the second approach is formulated as a system of linear equations. This work addresses one of the main challenges when dealing with variational quantum algorithms on general problems, namely designing an expressive and efficient quantum ansatz working on tracking events with fixed detector geometry. For this purpose, we employed a quantum architecture search method based on Monte Carlo Tree Search to design the quantum circuits for different problem sizes. We provide experimental results to test our approach on both formulations for different problem sizes in terms of performance and computational cost.

\end{abstract}
\section{Introduction}



In large-scale particle physics experiments, particles are collided at high energies and at high frequencies of 40 MHz, in order to study the fundamental forces of nature. In a single collision event, hundreds of new particles are simultaneously produced and traverse through sensitive detection layers where they deposit small amounts of energy, resulting in so-called \textit{hits} in the detectors. These hits are then reconstructed with fast algorithms into particle trajectories or \textit{tracks} that are used in subsequent analyses. In the upcoming High Luminosity phase of the Large Hadron Collider (LHC), the number of simultaneous collisions will increase significantly leading to the production of unprecedented data volumes to be processed. The increase in complexity presents a challenge still to be resolved, since the track reconstruction task scales to the power of 2-3 with the number of hits per layer.\\
\phantom{x}\hspace{1ex}Various approaches have been pursued \cite{di2024quantum}, at the time of writing the most performant approach is based on GPU parallel track reconstruction of events \cite{CAMPORAPEREZ2021101422SearchByTriplet}. There have been attempts to leverage quantum computing to solve the tracking problem. As tracking can be expressed as a Quadratic Unconstrained Binary Optimization (QUBO) problem many quantum algorithms can be applied. Some popular approaches are; Quantum Graph Neural Networks \cite{crippa2023quantumQGNN,LUXEqTrackQGNN,Tüysüz2021QGNN}, quantum annealing \cite{Bapst:2019llh,refId0Annealing}, quantum annealing-inspired algorithms \cite{okawa2024quantumannealinginspiredalgorithms} and Variational Quantum Eigensolver (VQE) using a sub-QUBO formulation of the problem and tested it on real hardware \cite{funcke2023studyingvqe}. 
Moreover, the tracking problem has been mapped to a linear system of equations and in \cite{nicotra2023quantum} the authors prove that it fulfills all the properties necessary to employ the Harrow–Hassidim–Lloyd (HHL) algorithm, which gives theoretical guarantees for an exponential advantage compared to classical techniques in terms of computational complexity. Although experimental results have shown promising results, a more extensive application of the HHL algorithm is still unfeasible on Near Intermediate-Scale Quantum (NISQ) devices \cite{preskill2018quantum}.  \\
\phantom{x}\hspace{1ex}
In this article, we employ the hardware-tailored approach of the Variational Quantum Algorithms (VQAs) for the particle tracking problem, that allows consideration of the user's available hardware specifications. We address one of the main limitations of these methods, that is the choice of the quantum circuit ansatz \cite{cerezo2021variational}, as also highlighted for our specific problem in \cite{funcke2023studyingvqe}. VQAs are hybrid quantum-classical algorithms where the problem of interest is encoded into an optimization task over parameterized quantum circuits. The choice of quantum circuit to optimize on, known as \textit{ansatz}, significantly affects the performance of the algorithm \cite{cerezo2021variational}. In this context, it the automatic design of ansatz for VQA emerged as a relevant research direction, known as \textit{quantum architecture search}, or also \textit{quantum ansatz search}. It can be seen as the quantum analogue of designing the structure on Artificial Neural Networks \cite{ren2021comprehensive}. For this scope, we employ an automated quantum ansatz search technique introduced in \cite{vincenzo} based on Monte Carlo Tree Search (MCTS) \cite{coulom2007monte}.
Moreover, we provide a deeper overview on the performance of Variational Quantum Algorithms \cite{cerezo2021variational} for particle track reconstruction by considering two different formulations. One is a ground state energy problem and is solved through a VQE \cite{kandala2017hardware}, similar to \cite{funcke2023studyingvqe}. While the other is encoded in a system of linear equations, equivalent to the HHL formulation \cite{nicotra2023} and solved through the Variational Quantum Linear Solver (VQLS) \cite{bravo2023variational}. Note that the ansatz search problem relates to both formulations. Hence, MCTS is used to design the ansatz for the VQE and VQLS independently and for all the tested problem sizes.


The goal is to find an ansatz that solves the tracking problem for different problem sizes and configurations. To test our approach we provide experimental results based on noiseless simulations of both problem formulations. 


\section{Particle Track Reconstruction} \label{Particle Track}

\begin{figure}[b!]
    \centering
    \includegraphics[scale=0.3]{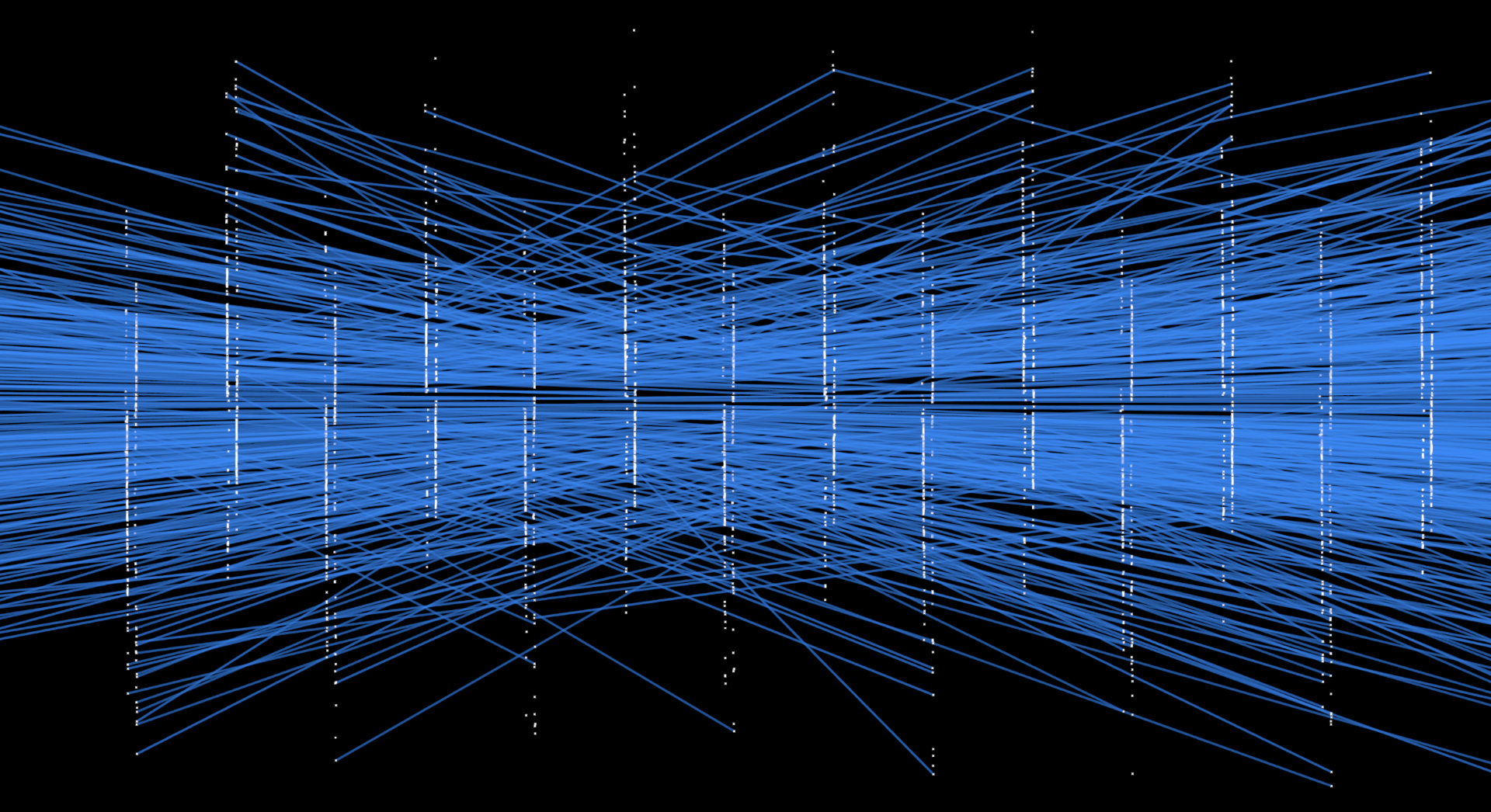}
    \caption{A full event example of the LHCb VELO detector. The white dots represent the detector hits and the blue lines are the reconstructed tracks. Credit: D. Nicotra.}
    \label{velo}
\end{figure}

In this paper, we address the particle tracking problem for a simplified version of the VELO subdetector of the LHCb experiment at the LHC at CERN \cite{The_LHCb_Collaboration_2008}. This is the subdetector immediately surrounding the particle collision point. This detector consists of two sides of 26 vertically oriented pixel sensors arranged along the beam line positioned around the LHC collision point. The detector geometry, particle hits and particle tracks can be seen in Figure \ref{velo}. This figure shows the complexity of a simulated event under a realistic detector response. One key aspect of the VELO is that due to the lack of a magnetic field, the particle tracks are straight lines, which can also be seen in Figure \ref{velo}. 

Solving the particle track reconstruction using quantum algorithms requires us to find a representation of the problem in such a way that it takes advantage of the quantum computation. Considering the detector geometry we construct a fully connected graph where the graph layers are the detector modules. Each layer has all the hits that pass through that detector layer and each connection represents a possible track segment (doublet). 

\begin{figure}[t!]
    \centering
    \includegraphics[scale=0.5]{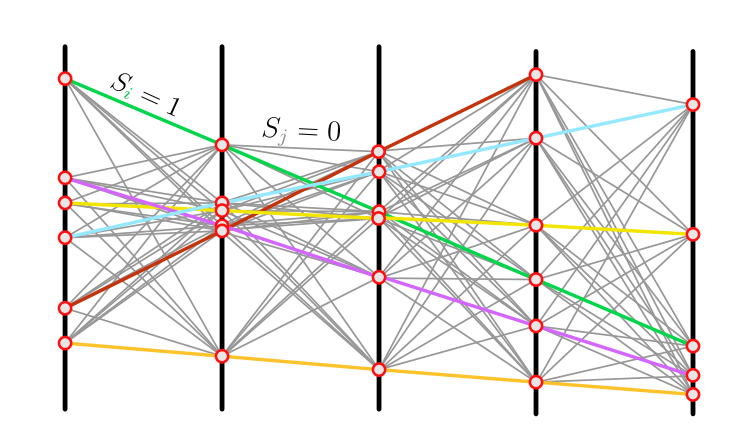}
    \caption{Graph representation of an event with 5 detector layers and 6 particles (Toy Model). The vertical lines represent the detection layers, and the red circles are the hits interconnected with doublets/segments. The full-color segments are part of the found tracks, the grey segments are combinatoric backgrounds. Where $S_i$ and $S_j$ are doublets with their value indicating if they are part of a track \cite{nicotra2023quantum}.}
    \label{event_graph}
\end{figure}

We assign a binary value to each track segment, $S_i = \{ 0, 1 \}$, where a $1$ indicates the doublet being part of a genuine particle track and a $0$ represents the contrary, this construction can be seen in Figure \ref{event_graph}. 
These doublets will be the basis of constructing an Ising-like Hamiltonian, described in Section \ref{variational_ptr}.   

To test the algorithms in a quantum simulator we use a toy model of the VELO detector, which was introduced in \cite{nicotra2023quantum} and is available publicly through the associated GitHub page \cite{nicotra2023}. This model allows us to test tracking scenarios where we can define the number of particles and the number of detector layers, a visual representation of this is Figure \ref{event_graph}, which shows an event. In this model, the segments/doublets are made by all possible connections between hits on adjacent layers, and tracks are a collection of aligned segments of length $layers-1$. 



\section{Variational Quantum Algorithms for Particle Track Reconstruction} \label{variational_ptr}
Variational Quantum Algorithms (VQAs) are hybrid quantum-classical algorithms where the problem of interest is encoded into an optimization task over parameterized quantum circuits \cite{cerezo2021variational}. 
Solving the particle track reconstruction problem with VQAs consists of three main parts. First encoding the classical combinatorial problem in a quantum setting for VQAs. Then, design a quantum ansatz that is expressible enough for different collision events with fixed detector geometry. Finally, optimize the angle parameters of the ansatz designed.
In Section \ref{vqe}, we describe the Variational Quantum Eigensolver (VQE) used to find the ground state energy of the Hamiltonian introduced in Eq. \eqref{H_vqe} \cite{kandala2017hardware}. In Section \ref{vqls}, we describe the Variational Quantum Linear Solver (VQLS) \cite{bravo2023variational} used to solve the system of linear equations defined in Eq. \eqref{H_vqls}. In Section \ref{encoding}, we describe the key encoding process of the classical problem in a circuit-based quantum computing setup.

\subsection{Variational Quantum Eigensolver} \label{vqe}
The Variational Quantum Eigensolver \cite{kandala2017hardware} has advanced quantum chemistry by enabling the study of molecules on NISQ devices. VQE defines the problem through the system's physical properties, such as its geometric structure and atomic correlations, which are described by the system's Hamiltonian. Similarly, the geometry of straight particle trajectories is encoded in the Hamiltonian in our application.
Then, the objective is to minimize the cost function $C$, or energy, defined by the expectation value of the Hamiltonian $H$ of the system:
\begin{equation}
    C_{VQE}(\theta) = \bra{\psi(\theta)} H_{VQE} \ket{\psi(\theta)}
\end{equation}
where $\psi$ is the variational circuit ansatz and $\theta$ its gate parameters. The minimum value $\Bar{C}_{VQE}$ corresponds to the ground state energy
\begin{equation}
    \Bar{C}_{VQE} = \min_\psi \bra{\psi} H_{VQE} \ket{\psi}
\end{equation}
corresponding to the ground state $\Bar{\psi}$ composed of both the optimal circuit structure $\Bar{\psi}$ and gate parameters. The rounding of the amplitudes of the ground state $\Bar{\psi}(\theta)$ to a binary vector is the solution to the tracking problem.

\subsubsection{$\mathcal{H}_{VQE}$:}
To solve the tracking problem using a VQE we define an Ising-like Hamiltonian whose ground state eigenvector represents the solution to the particle tracking problem. This is equivalent to minimizing the energy function of a Hopfield network \cite{Hopfield1982-gk}. In this paper, we use a slightly modified version of the Denby-Peterson (DP) Hamiltonian \cite{DENBY1988429,PETERSON1989537}, which has been previously used to solve tracking problems\cite{Zlokapa_2021DP_hamiltonian}. We define this Hamiltonian as

\begin{equation} 
    \mathcal{H}_{VQE} = \mathcal{H}_{ang}(\mathbf{S},\epsilon) + \gamma \mathcal{H}_{bif}(\mathbf{S}) + \delta \mathcal{H}_{occ}(\mathbf{S},N_{hits}) 
    \label{H_vqe}
\end{equation}

$\mathcal{H}_{ang}$ is a modified angular term defined in \cite{nicotra2023quantum}, where $\mathbf{S}$ is a doublet as defined in Figure \ref{event_graph}. It creates a step function that promotes the construction of long smooth tracks and forbids forming strongly kinked trajectories. The parameter $\epsilon$ represents the threshold that defines aligned segments. The $\mathcal{H}_{bif}$ penalizes bifurcations, where a single track splits into two or more branches, which are usually non-physical and tend to represent noise rather than real particle tracks. $\mathcal{H}_{occ}$ constrains the number of reconstructed segments to be approximately equivalent to the number of hits in the detector, ensuring that each hit contributes to the reconstruction and preventing an overabundance of segments. The parameters $\gamma$ and $\delta$ allow for the fine-tuning of the Hamiltonian to adapt to different detector or data characteristics.

\subsection{Variational Quantum Linear Solver}\label{vqls}
The Variational Quantum Linear Solver (VQLS) \cite{bravo2023variational} leverages variational methods to solve linear systems of equations by optimizing a cost function. A system of linear equations is defined by a matrix $A$ and vector $b$, where the goal is to find a vector $x$ such that:

\begin{equation}
    A x = b
\end{equation}

Given finding vector $x$ is our algorithms objective, our algorithms inputs are $A$, $b$ and ansatz $V(\theta)$. The matrix $A$ and vector $b$ need to be efficiently encoded into the quantum system, this process is described in Section \ref{encoding}. We implement the linear map $A$ as a coherent probabilistic operation \cite{bravo2023variational}. The VQLS aims to prepare the state $\ket{\psi}$ where $x \approx \ket{\psi}$, such that $A\ket{\psi}$ is proportional to $\ket{b}$

\begin{equation}
    \ket{\phi} = A\ket{\psi} \approx \ket{b}
\end{equation}

To approximate the solution $\ket{\phi}$, we apply a problem-specific ansatz $V(\theta)$ to the ground state $\ket{0}$, resulting in the state $\ket{\psi} = V(\theta)\ket{0}$. The goal is to optimize the parameters $\theta$ to maximize the overlap between the quantum states $\ket{\phi}$ and $\ket{b}$. This overlap is quantified by the following cost function:

\begin{equation}
    C_{VQLS}(\theta) = 1 - |\braket{b|\phi}|^2
\end{equation}

The cost function is an implementation of the intuitive thought that given a $\ket{\phi}$ and projector $\ket{\phi}\bra{\phi}$, we want a cost function based on the overlap between this projector and the subspace orthogonal to $\ket{b}$ \cite{bravo2023variational}. However, we do not implement $A\ket{\psi}$ directly we rather measure the ratio between the probability of our circuit being in the ground state and the probability of the ancilla qubits being in the ground state. Taken as an indirect measurement of the overlap, in what \cite{pennylane_vqls_2024} names the Coherent VQLS, which defines the overlap as

\begin{equation}
    |\braket{b|\phi}|^2 = \frac{P(All \: qubits \: are \: in \: the \: ground \: state)}{P(Ancilla \: qubits \: are \: in \: the \: ground \: state)}
\end{equation}

The cost function can be computed using easily evaluated quantities: the ground state vector, the full quantum system state, and the ancilla qubit states.

\subsubsection{$\mathcal{H}_{VQLS}$:}
For the VQLS algorithm, we need to construct a system of linear equations whose solution solves the tracking problem. In \cite{nicotra2023quantum} it is shown that we can simplify the VQE Hamiltonian maintaining the angular term only and apply a relaxation procedure to make the Hamiltonian differentiable. Then by taking the differential of the Hamiltonian, it can be shown that

\begin{equation}
   \nabla \mathcal{H} = - A \mathbf{S} + b
   \label{H_diff}
\end{equation}

Therefore finding the minimum of the Hamiltonian $\nabla \mathcal{H} = 0$ is equivalent to solving $A \mathbf{S} = b$. The VQLS Hamiltonian is defined as
\begin{equation}
    \mathcal{H}_{VQLS} = \mathcal{H}_{ang}(\mathbf{S},\epsilon) + \zeta \mathcal{H}_{spec}(\mathbf{S}) + \eta \mathcal{H}_{gap}(\mathbf{S})
    \label{H_vqls}
\end{equation}

The spectral term $\mathcal{H}_{spec}$ is included to ensure that $\mathcal{H}_{VQLS}$ is a positive-definite matrix and shifts the value of the diagonal of $A$ by a constant $\zeta$. The $\mathcal{H}_{gap}$ ensures that there is a gap in the solution spectrum of $\mathbf{S}$, which defines the threshold that discriminate between solution and non-solution segments, by contributing to the matrix $A$ and defining $b$. 

\begin{equation}
    b = \eta(1,1,...,1)
    \label{vec_b}
\end{equation}

\subsection{Encoding}\label{encoding}
The encoding of classical information into a quantum system is a key aspect of any quantum computing algorithm, as to leverage the power of the quantum computation we must first be able to efficiently encode our classical information. We implement two encoding protocols for our VQE and VQLS problems. For both cases, we first construct matrices $A_{VQE}$ and $A_{VQLS}$ from their respective Hamiltonians defined in Eq. \ref{H_vqe} and \ref{H_vqls}. We then take these matrices and perform a Pauli decomposition, which describes our matrices as linear combinations of tensor products of the Pauli matrices $\sigma_i$, where $i \in \{0, 1, 2, 3 \}$ \setcounter{footnote}{0} \footnote{The Pauli matrices $\sigma_i$ form a complete basis spanning the space of all $2 \times 2$ matrices:
\[
\sigma_0 = \begin{pmatrix} 
1 & 0 \\ 
0 & 1 
\end{pmatrix}, \quad
\sigma_1 = \begin{pmatrix} 
0 & 1 \\ 
1 & 0 
\end{pmatrix}, \quad
\sigma_2 = \begin{pmatrix} 
0 & -i \\ 
i & 0 
\end{pmatrix}, \quad
\sigma_3 = \begin{pmatrix} 
1 & 0 \\ 
0 & -1 
\end{pmatrix}.
\]
}. The Pauli decomposition can be expressed as



\begin{equation}
    A = \sum^{n}_{i_1, i_2,...,i_n} c_{i_1, i_2,...,i_n} \sigma_{i_1} \otimes \sigma_{i_2} \otimes ... \otimes \sigma_{i_n}
\end{equation}
where $
    c_{i_1, i_2,...,i_n} = \frac{1}{2^n} \mathbf{Tr}(A \cdot \sigma_{i_1} \otimes \sigma_{i_2} \otimes ... \otimes \sigma_{i_n})$.

We use the Pennylane implementation of the Pauli decomposition \cite{hamaguchi2024pennylanepauli}, which has a worst-case time complexity of $\mathcal{O}(n4^n)$ where $n$ is the number of qubits needed to encode the matrix. For the VQLS we require one additional encoding step which is that of the vector $b$ defined in Eq. \ref{vec_b}. In general, one needs to construct a unitary $U_b$ such that $\ket{b} = U_b \ket{0}$, but since our $b$ is trivial as we set $\eta = 1$ in Eq. \ref{vec_b}, therefore all we need to do to encode it is apply a Hadamard gate to each qubit encoding $b$ \footnote{The Hadamard gate is a one-qubit operation that maps the classical computational basis states, $\ket{0}$ and $\ket{1}$, to equal superpositions of both. Its matrix representation is:
\[
H = \frac{1}{\sqrt{2}} \begin{pmatrix}
1 & 1 \\
1 & -1
\end{pmatrix}.
\]
}.

For the VQLS Pauli decomposition, the following relations hold; $\sum^{L-1}_{l=0} c_l = 1 $ and $ c_l \geq 0$, which constructs a normalized probability distribution where $l$ is padded to a power of $2$, where $L$ is the number of Pauli gates in the decomposition. We need $\log_2 l$ ancilla qubits to encode the matrix $A$. This allows us to carry out the following steps: 

\begin{enumerate}
    \item $\mathbf{Pauli \: Matrix \: Encoding}$: We apply multi-controlled $\sigma_i$ for each index indicated by the Pauli decomposition. Each $A_l$ is used as a multi-controlled gate, where the Pauli decomposition gives the target qubit index \cite{hamaguchi2024pennylanepauli}, and the control qubits are defined by the binary representation of the $l$ index of matrix $A_l$.
    \item $\mathbf{Coefficient \: Encoding}$: Next we embed the amplitudes of the coefficients into the state of the ancilla qubit. This is done through a nonlinear feature map that maps the coefficients to a quantum Hilbert space \cite{Schuld_2019AE}. Pennylane $\mathbf{AmplitudeEmbedding}$ functionality is used to embed our coefficients \cite{bergholm2022pennylaneWhitePaper}.
\end{enumerate}

\section{Quantum Ansatz Search} \label{ansatz search}
In the context of VQAs, an \textit{ansatz} refers to the "guess" on the parameterized quantum circuit whose parameters have to be optimized in order to minimize the cost function. The performance of VQAs are significantly dependent on the ansatz choice, which defines their expressability and trainability \cite{cerezo2021variational}. An ansatz consists of an ordered sequence of quantum gates $V$ with their respective positions and corresponding angle parameters $\theta$. 
When the problem under study presents peculiar symmetries it is possible to define guidelines to design `problem-inspired' ansatz \cite{kandala2017hardware,lee2018generalized,fedorov2022vqe}. In our problem,, we have not extensively searched for symmetries to exploit in order to solve the problem for different events and different problem sizes. However, it might be a fruitful research direction. In fact, using brute-force algorithms to find a suitable ansatz is unfeasible given the exponential scaling of the computational time with respect to the problem size \cite{nielsen2010quantum}. In this scenario, Quantum Architecture Search (QAS), also known as quantum ansatz search, emerges as a crucial paradigm to automatically explore the architectures of parameterized quantum circuits by leveraging computational resources \cite{kandala2017hardware,grimsley2019adaptive,tang2021qubit,zhu2022adaptive}. 

In the following sections, we describe the ansatz search problem in the context of particle track reconstruction. Note that we have to search for a suitable ansatz for both problem formulations independently. In both cases we tackle the problem through the Monte Carlo Tree Search introduced in \cite{vincenzo}. 

\subsection{Ansatz Search for Particle Tracking}
The quantum ansatz problem consists of finding a suitable structure for the variational quantum circuit, that given the geometry of the detector, solves particle track reconstruction.
\begin{figure}[t!]
    \centering
    \includegraphics[scale=0.15]{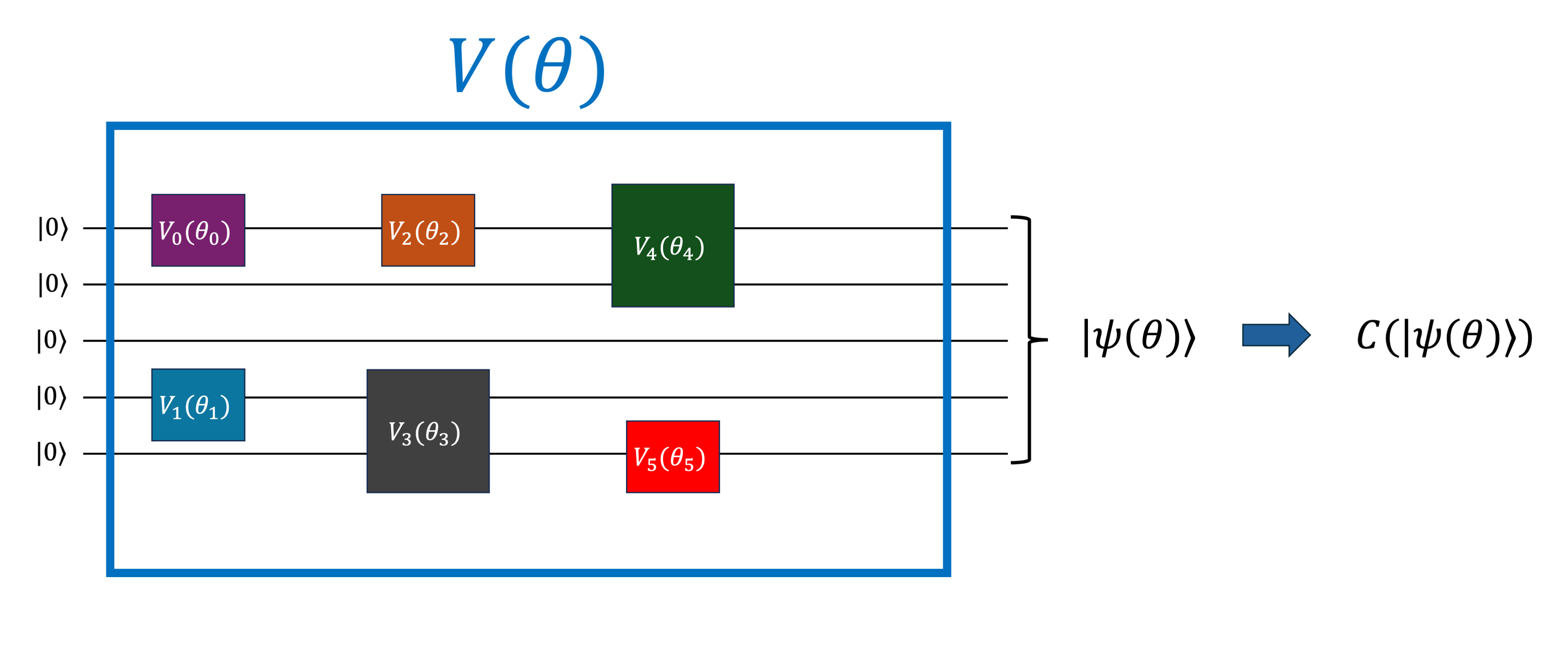}
    \caption{The goal in QAS is to define the ordered sequence of quantum gates $V_i(\theta_i)$ composing the variational state $V(\theta)\ket{0}$, which optimizes a given objective function $C$.}
    \label{qas_scheme}
\end{figure}
The variational quantum state $\ket{\psi}= V\ket{0}$ can be generally written as a sequence of parameterized and non-parameterized unitaries
\begin{equation}
    V (\theta)= \prod_{i=1}^l V_i (\theta_i)  \, ,\label{variational_circuit}
\end{equation}
where $l$ is the number of quantum gates $V_i$ chosen from the universal gate set $G=( CX, R_x, R_y, R_z)$ consisting of the controlled-NOT gate and of the parameterized single-qubit rotations \cite{nielsen2010quantum} and $\theta =(\theta_0,\theta_1, ..., \theta_l) \in \mathbb{R}^{l}$ is the vector of angle parameters. In this notation, the $\theta_i$ is zero-dimensional if $V_i$ is a non-parameterized gate, the starting state is $\ket{0}=\ket{0}^{\otimes n }$, $V$ is an $n$-qubit operator and $ V_i$ is either a single-qubit or $2$-qubit operator tensored with the identity operator on the subspace of the unaffected qubit(s).

\subsection{Design Variational Quantum Ansatz through Monte Carlo Tree Search} \label{mcts}
Monte Carlo Tree (MCTS) is a best-first search method whose basic implementation does not require any domain-specific knowledge. This trait is particularly useful for the design of algorithms in artificial general intelligence \cite{finnsson2012generalized,sironi2018self}. 

MCTS is based on a randomized exploration of the search space. Using the results of previous explorations, MCTS gradually builds up a search tree in memory, and successively becomes better at accurately estimating the values of the most promising actions. It consists of four strategic steps \cite{chaslot}, repeated as long as there is time left. (1) In the \textit{selection} step the tree is traversed from the root node downwards until a state is chosen, which has not been stored in the tree. (2) Next, in the \textit{roll-out} step, actions are randomly chosen until a terminal state is reached. (3) Subsequently, in the \textit{expansion} step one or more states encountered along the roll-out are added to the tree. (4) Finally, in the \textit{backpropagation} step, the reward is propagated back along the previously traversed path up to the root node, where node statistics are updated accordingly. 
MCTS grows its search tree gradually by executing the four steps described above. Such an iteration is called a full simulation. 
The selection algorithm used in this work is the UCB1 \cite{auer2002finite}, which balances exploitation and exploration of the search \cite{browne2012survey}. 
Given a state $s$ and the set of all possible actions $\mathcal{A}$, our MCTS takes the action $a^*$ with the highest $UCB$ value
\begin{equation}
a^* = \arg \max_{a \in \mathcal{A}} \, UCB(s, a)   
\end{equation}
\begin{equation}
UCB(s, a) = \frac{Q_{(s,a)}}{N_{(s,a)}}+ c\, \sqrt{\frac{\log N_s}{N_{(s,a)}}} \label{uct}
\end{equation}
where $N_{(s,a)}$ is the number of times the agent took the action $a$ from the state $s$, $N_s = \sum_{a\in \mathcal{A}} N_{(s,a)}$ is the total number of times the agent visited the state $s$ and $Q_{(s,a)}$ is the cumulative reward the agent gained by taking the action $a$ from the state $s$. The constant $c$ is a parameter that controls the degree of exploration, captured by the second term of Eq.~\ref{uct} versus the exploitation captured by the first term. Here, $c$ has been set to $0.4$ based on previous fine-tuning experiments.

The implementation of MCTS employed to design the ansatz is an agnostic approach using a random roll-out \cite{vincenzo}. Here each node corresponds to a $n$-qubit quantum circuit and each move corresponds to a specific modification of it. 

The reward for MCTS is a real-valued function on the domain of $n$-qubit quantum circuits defined as follows for the two independent problems. The reward $R_{VQE}$ for VQE is  
\begin{equation}
    R_{VQE} (\ket{\psi (\theta)}) = -\bra{\psi(\theta)}H\ket{\psi(\theta)}.
\end{equation}
the reward $R_{VQLS}$ the variational quantum linear solver is
\begin{equation}
    R_{VQLS}(\ket{\psi (\theta)}) = \exp(-C_{VQLS}(\theta)).
\end{equation}
Note that the reward function is well-defined on all the tree nodes. Since each of them can be considered a potential terminal state, the roll-out step is not necessary to evaluate a new visited node but it may give insights into future moves during the search.

The search starts from the quantum circuit with a Hadamard gate applied on each qubit, previously initialized to $\ket{0}$. This allows to start from a non-classical state, given by the equal superposition of all computational basis states.
MCTS explores the search space by sampling from four classes of allowed actions:

\begin{enumerate}
    \item \textit{adding} (A) a random gate on a random qubit at the end of the circuit;
    \item \textit{swapping} (S) a random gate in the circuit with a new one;
    \item \textit{deleting} (D) from the circuit a gate at random position;
    \item \textit{changing} (C) the angle parameter $\theta_i$ of a randomly chosen parameterized gate in $\theta_i+\epsilon$, where $\epsilon \sim \mathcal{N}(0, \Delta \theta)$ is sampled from a normal distribution with mean zero and standard deviation $\Delta \theta$.
\end{enumerate}

\begin{figure}[!b]
    \centering
    \includegraphics[scale=0.5]{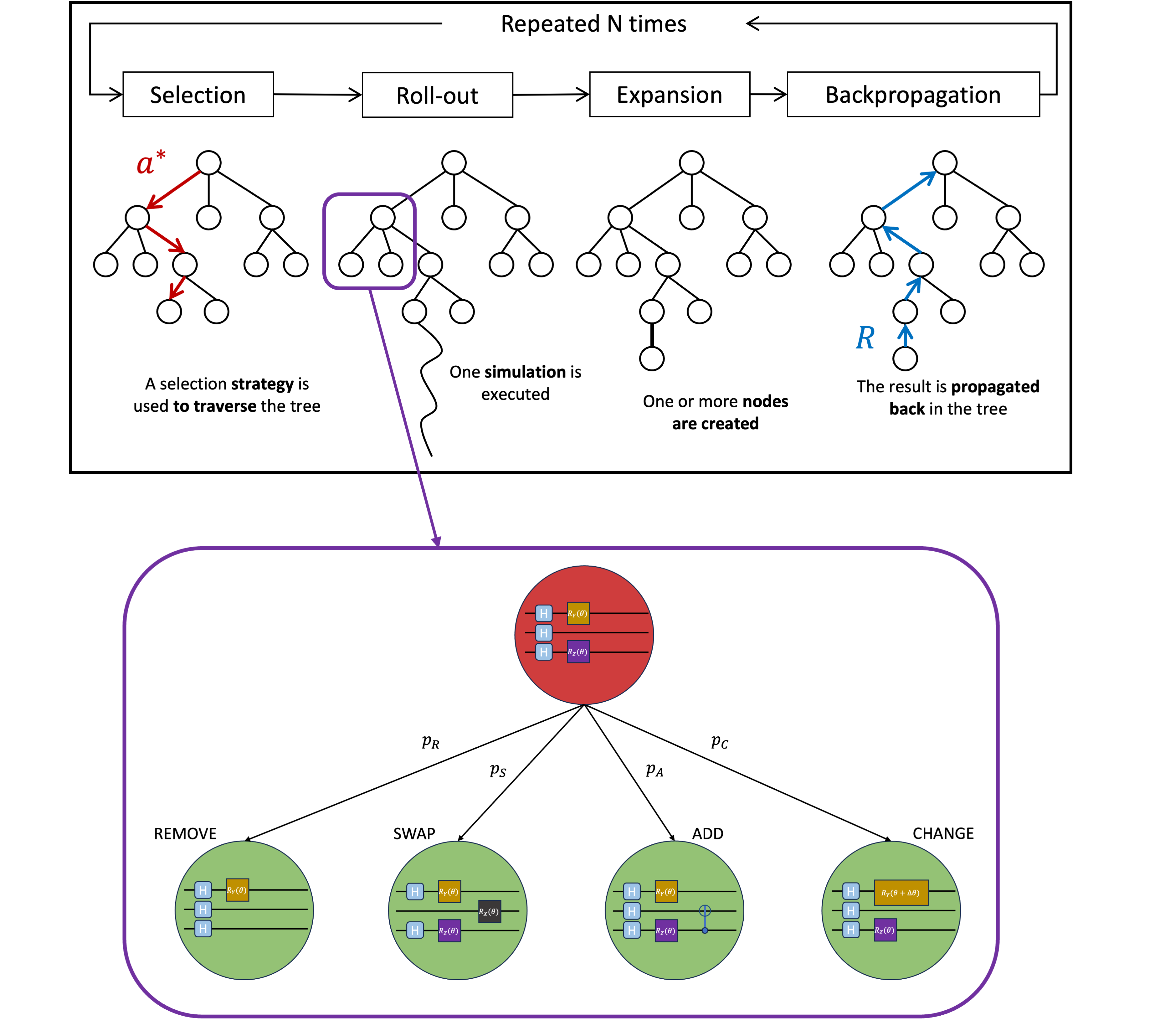}
    \caption{Monte Carlo Tree Search scheme for quantum circuit design, taken from \cite{vincenzo}. In our QAS framework, the action space is defined by sampling a discrete number of quantum circuit modifications from a continuous set of them.}
    \label{mcts_scheme}
\end{figure}

MCTS chooses between the four classes of actions by sampling from a probability mass distribution $p=(p_A, p_S, p_C, p_D)$,
where $p_A$, $p_S$, $p_C$, $p_D$ correspond to the probabilities of choosing the respective action. A clarifying scheme is represented in Figure \ref{mcts_scheme}. The hyperparameters $p$ and $\Delta \theta$ have been fixed according to the previous work \cite{vincenzo}. This technique is inspired by a framework proposed by Franken et al. \cite{franken2022quantum} for an evolutionary strategy on quantum circuits. The quantum gates are sampled from the universal gate set $G$, introduced in the previous section. 
A progressive widening technique \cite{chaslot} is also incorporated in the MCTS to restrict the infinite discrete number of moves allowed from a state $s$ to a finite number of moves:
\begin{equation}
    k_s=\lceil \beta_{PW} N_s^{\alpha_{PW}} \rceil    \label{progressive_widening}
\end{equation}
where the hyperparameter $\alpha_{PW} \in ]0,1[$ and $\beta_{PW} >0$ have been fixed according to the previous work: $\alpha_{PW}=0.3$ and $\beta_{PW}=1$ \cite{vincenzo}.
Since we are interested in the whole sequence of gates an action-by-action search is employed in order to distribute the search time along all the levels of the tree \cite{baier2012nested,schadd2012single}. Once a level of the tree is sufficiently explored, the MCTS commits to the best action. 

At the end of the search, the \textit{best path} that builds the ansatz is retrieved according to the method described in \cite{vincenzo}. 

\subsection{Parameter Optimization} 
The last step is fine-tuning the angle parameters of the ansatz found by the MCTS. It is carried out by the ADAM optimizer \cite{kingma2014adam}, a gradient-based classical optimizer, chosen for its robustness and adaptability. It adjusts the learning rates adaptively for each parameter by computing the first and second moments of the gradients. To apply a gradient descent technique in the circuit-based quantum computing setting we used the parameter-shift rule \cite{wierichs2022general}.

\section{Results}\label{experiments}
We explored different problem sizes with increasing combinatorial complexity to evaluate the performance of the MCTS, introduced in Ref \cite{vincenzo} to design quantum ansatz for the VQE and VQLS formulations of the particle track reconstruction problem.. Then, we construct $H_{VQE}$ and $H_{VQLS}$ for the toy model of the VELO detector geometry, introduced in 
Section \ref{Particle Track}. Using this model we can specify the number of particles and layers. From this, we derive the size $N\times N$ of the hamiltonian, where $N=(\text{n° of particles})^2 \times (\text{n° of layers}-1)$. The number of qubits required in both formulations is $ n=\log_2N$ and $N$ grows logarithmically with the problem size. We refer the reader to Table \ref{table_problem}. 

\begin{table}[!t]
    \centering
    \caption{Generation of the particle track reconstruction with a toy model \cite{nicotra2023quantum}.}
    \label{table_problem}
    \begin{tabular}{
        @{} >{\centering\arraybackslash}m{1.3cm}  
        >{\centering\arraybackslash}m{3.1cm}  
        >{\centering\arraybackslash}m{1.8cm}      
        >{\centering\arraybackslash}m{1.5cm}
        >{\centering\arraybackslash}m{2.0cm} 
        >{\centering\arraybackslash}m{2.0cm}     
        @{}
    }
        \toprule
        \textbf{Qubits} $n$&\textbf{Hamiltonian Size} $2^n\times 2^n$& \textbf{Particles} & \textbf{Layers} & \textbf{Solution Segments} & \textbf{Total Segments} \\
        \midrule
        3&$8\times 8$   & 2 & 3 & 4 &8 \\
        4&$16\times 16$ & 2 & 5 & 8 & 16 \\
        5&$32\times 32$ & 4 & 3 & 8 &32\\
        6&$64\times 64$ & 4 & 5 & 16 &64\\
        7&$128\times 128$ & 8 & 3  & 16 &128\\
        8&$256\times 256$ & 8 & 5  & 32 &256\\
        \bottomrule
    \end{tabular}
    \label{table 1}
\end{table}
 MCTS designs circuits tailored on the quantum hardware available to the user \cite{vincenzo}. In our case as we run simulated experiments, we fix a maximum depth to keep the running time low, we chose $50$. The MCTS has been equipped with a computational budget fixed to $10^4$ for problems with $n\leq 5 $, while it has been fixed to $10^5$ for bigger $n$ as the search space to explore is larger. 

\subsection{Comparison VQE-VQLS}
\begin{figure}[t!]
    \centering
    \includegraphics[scale=0.4]{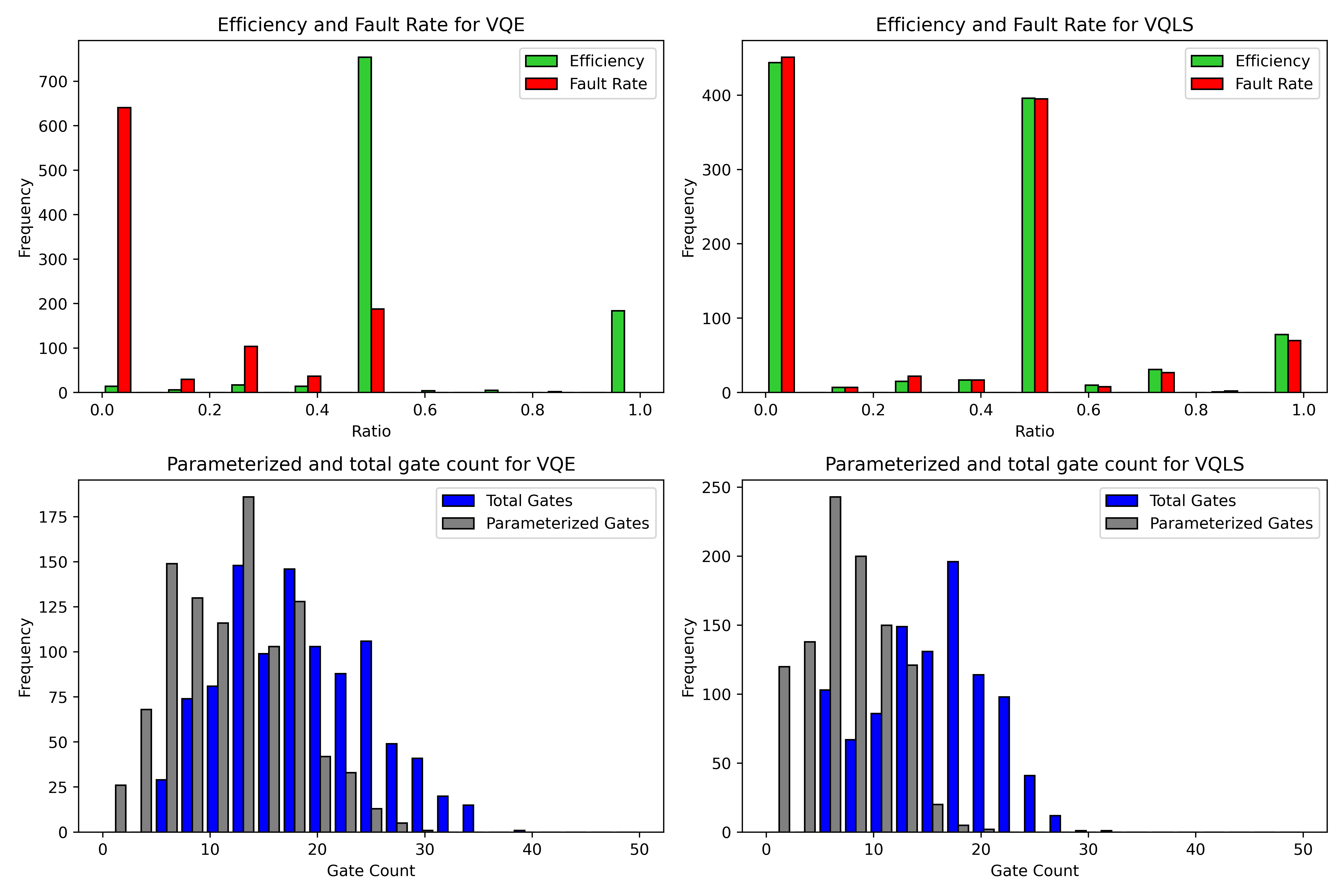}
    \caption{Experimental results for the problem with $n=4$ qubits .}
    \label{fig:16x16 distribution}
\end{figure}

The quality of an ansatz is evaluated on its ability to generate correct tracking solutions over $1000$ independent runs.
We define the \textit{efficiency} as the ratio between the doublets predicted correctly and the total number of doublets in the tracks. The \textit{fault rate} is defined as the ratio between the number of doublets that are mistakenly predicted as part of tracks over the total number of segments that do not belong to any track.
The numerical results that compare VQE and VQLS for $n=4$ are plotted in Figure \ref{fig:16x16 distribution}. The complementary results for the other problem size are summarized in Table \ref{table_vqe} where we averaged the efficiency and fault rate over the $1000$ independent runs.

\begin{table}[!b]
    \centering
    \caption{VQE and VQLS results as mean values and standard deviation over $1000$ independent runs. VQLS was limited to $n=6$ due to high computational cost and memory requirements.}
    \label{table_vqe}
    \begin{tabular}{
        @{} >{\centering\arraybackslash}m{2.8cm}  
        >{\centering\arraybackslash}m{2cm}      
        >{\centering\arraybackslash}m{2cm}      
        >{\centering\arraybackslash}m{2cm}      
        >{\centering\arraybackslash}m{2.8cm}      
        @{}
    }
        \toprule
        \textbf{Hamiltonian Size $2^n\times 2^n$} & \textbf{Efficiency} & \textbf{Fault Rate} & \textbf{Total Gates} & \textbf{Parameterized Gates} \\
        \midrule
        \multicolumn{5}{c}{\textbf{VQE}} \\
        \midrule
        $8\times 8$   & 0.56 ($\pm$0.18) & 0.39 ($\pm$0.20) & 16 ($\pm$6) & 12 ($\pm$6) \\
        $16\times 16$ & 0.58 ($\pm$0.21) & 0.14 ($\pm$0.20) & 18 ($\pm$7) & 12 ($\pm$6) \\
        $32\times 32$ & 0.42 ($\pm$0.17) & 0.17 ($\pm$0.07) & 20 ($\pm$5) & 14 ($\pm$5) \\
        $64\times 64$ & 0.51 ($\pm$0.19) & 0.14 ($\pm$0.08) & 25 ($\pm$6) & 17 ($\pm $5) \\
        $128\times 128$ & 0.30 ($\pm$0.14) & 0.09 ($\pm$ 0.04)& 25 ($\pm$4) &  17 ($\pm $4) \\
        $256\times 256$ & 0.27 ($\pm$0.17) & 0.05 ($\pm$0.02) & 31 ($\pm$7) &  20 ($\pm $6) \\
        \midrule
        \multicolumn{5}{c}{\textbf{VQLS}} \\
        \midrule
        $8\times 8$   & 0.22 ($\pm$0.32) & 0.22 ($\pm$0.32) & 14 ($\pm$5) & 7 ($\pm$4) \\
        $16\times 16$ & 0.32 ($\pm$0.32) & 0.31 ($\pm$0.31) & 16 ($\pm$6) & 7 ($\pm$4) \\
        $32\times 32$ & 0.26 ($\pm$0.28) & 0.26 ($\pm$0.25) & 16 ($\pm$6) & 7 ($\pm$4) \\
        $64\times 64$ & 0.26 ($\pm$0.25) & 0.26 ($\pm$0.23) & 17 ($\pm$6) & 7 ($\pm $4) \\
        \bottomrule
    \end{tabular}
    \label{table 2}
\end{table}

In the both setups, some of the MCTS solutions are able to perfectly solve the tracking problem at multiple sizes, while many others are able to identify only a fraction of the tracks. Although the fixed computational budget given to MCTS and the constraint to produce quote shallow quantum circuit, it recognizes the solutions for different and increasing problem sizes. These perfect solutions can be seen in Figure \ref{fig:16x16 distribution}. Future work will study the properties of those circuits and pass the information directly to the model or filtering them with post-processing. 
The experiments revealed that VQE results in a more convenient formulation for MCTS compared to VQLS, on equal quantum circuit size and classical computational budget, see Figure \ref{fig:16x16 distribution} and Table \ref{table 2}. Note that the depth of quantum circuits can give serious implications on solution quality. In fact, the HHL algorithm requires much deeper circuits to "exactly" solve the same problem tackled by VQLS. While the HHL algorithm provides guarantees on the quality of the solution \cite{nicotra2023quantum}, the VQLS allows us to study larger problem sizes on NISQ devices, albeit with reduced performance.


\section{Conclusions}
In conclusion, our study tested the MCTS  approach introduced in \cite{vincenzo} to design the quantum ansatz for the Variational Quantum Eigensolver \cite{kandala2017hardware} and for the Variational Quantum Linear Solver \cite{bravo2023variational} for solving particle track reconstruction problem. 
Designing an ansatz for the VQE formulation turned out to be easier than for VQLS. We showed that MCTS manages to design rather shallow quantum circuits that provide suitable solutions for the toy model considered. However, as the problem size increases, the amount of classical computational resources needed to explore the exponentially larger search space of the quantum circuit is still quite high to compete with the classical state of the art. Employing an agnostic algorithm as the MCTS with random rollout \cite{vincenzo} limits the performance compared to more domain-specific approaches. The next step for this work is to integrate the intricacies of the problem domain into MCTS. It may be realized through a simulation strategy in the rollout step of the MCTS. Simulation strategies based on heuristics have the potential to significantly improve the performance of MCTS \cite{winands2019monte}. In case domain knowledge is not readily available, statistical methods can be employed to enhance the search. An example is the n-gram technique
, which has been successfully used in MCTS to detect promising action sequences \cite{tak2012n}.

In future work, we aim to evaluate the model in a more realistic scenario. On the one hand, expand the noise-less simulations with simulated noise or real-hardware experiments. On the other hand, address current limitations related to the toy model employed. Testing Hamiltonians corresponding to smaller events, padded with zeros, against ansatz configurations designed for larger events, will also facilitate a more realistic assessment of the MCTS scalability, reflecting practical scenarios where the number of particles is unknown a priori.

\begin{credits}
\subsubsection{Data Availability}
The data and code of this study are openly available on \href{https://github.com/VincenzoLipardi/MCTS-TrackReconstruction}{GitHub}.
\subsubsection{\ackname} 
We thank Davide Nicotra for the image in Figure \ref{velo} and Miriam Lucio Martinez for insightful comments on the first draft. This publication is part of the project Fast sensors and Algorithms for Space-time Tracking and Event Reconstruction (FASTER) with project number OCENW.XL21.XL21.076 of the research programme ENW - XL which is (partly) financed by the Dutch Research Council (NWO).

\subsubsection{\discintname}
The authors have no competing interests to declare that are relevant to the content of this article. 
\end{credits}

\bibliographystyle{splncs04}
\bibliography{bibliography}
\end{document}